\documentclass[A4paper,unsortedaddress,superscriptaddress,prl,twocolumn,showpacs]{revtex4}
\usepackage{graphicx}
\usepackage{amsmath}
\usepackage[T1]{fontenc}
\usepackage{color}

\begin{document}
\DeclareGraphicsExtensions{.pdf}

\title{Graphene on Ir(111): Physisorption with chemical modulation}

\author{Carsten Busse} 
\email{busse@ph2.uni-koeln.de}
\affiliation{II. Physikalisches Institut, Universit\"at zu K\"oln, 50937 K\"oln, Germany}
\author{Predrag Lazi\'{c}}
\affiliation{Peter Gr{\"u}nberg Institut (PGI) and Institute for Advanced Simulation (IAS), Forschungszentrum J{\"u}lich and JARA, 52425 J{\"u}lich, Germany}
\author{Rabie Djemour}
\affiliation{II. Physikalisches Institut, Universit\"at zu K\"oln, 50937 K\"oln, Germany}
\author{Johann Coraux}
\affiliation{Institut N\'{e}el, CNRS-UJF, 25 rue des Martyrs, BP 166, 38042 Grenoble cedex 9, France}
\author{Timm Gerber}
\affiliation{II. Physikalisches Institut, Universit\"at zu K\"oln, 50937 K\"oln, Germany}
\author{Nicolae Atodiresei}
\affiliation{Peter Gr{\"u}nberg Institut (PGI) and Institute for Advanced Simulation (IAS), Forschungszentrum J{\"u}lich and JARA, 52425 J{\"u}lich, Germany}
\author{Vasile Caciuc}
\affiliation{Peter Gr{\"u}nberg Institut (PGI) and Institute for Advanced Simulation (IAS), Forschungszentrum J{\"u}lich and JARA, 52425 J{\"u}lich, Germany}
\author{Radovan Brako}
\affiliation{Ru{\dj}er Bo{\v{s}}kovi{\'{c}} Institute, 10000 Zagreb, Croatia}
\author{Alpha T. N'Diaye}
\affiliation{II. Physikalisches Institut, Universit\"at zu K\"oln, 50937 K\"oln, Germany}
\author{Stefan Bl{\"u}gel}
\affiliation{Peter Gr{\"u}nberg Institut (PGI) and Institute for Advanced Simulation (IAS), Forschungszentrum J{\"u}lich and JARA, 52425 J{\"u}lich, Germany}
\author{J{\"o}rg Zegenhagen}\affiliation{European Synchrotron Radiation Facility (ESRF), 38043 Grenoble cedex 9, France}
\author{Thomas Michely}
\affiliation{II. Physikalisches Institut, Universit\"at zu K\"oln, 50937 K\"oln, Germany}

\date{\today}

\begin{abstract}
The non-local van der Waals density functional (vdW-DF) approach is applied to calculate the binding of graphene to Ir(111). The precise agreement of the calculated mean height $\bar{h}=3.41$~{\AA} of the C atoms with their mean height $\bar{h}=(3.38 \pm 0.04)$~{\AA} as measured by the X-ray standing wave (XSW) technique provides a benchmark for the applicability of the non-local functional. We find bonding of graphene to Ir(111) to be due to the van der Waals interaction with an antibonding average contribution from chemical interaction. Despite its globally repulsive character, in certain areas of the large graphene moiré unit cell charge accumulation between Ir substrate and graphene C atoms is observed, signaling a weak covalent bond formation.

\end{abstract}


\maketitle

Epitaxial growth on metals is a key method to produce high quality graphene on large scales \cite{Coraux2008,Martoccia2008}. Owing to the strength of the C-C bonds, large (incommensurate or weakly commensurate) superstructures are found for lattice-mismatched systems. The extend of superperiodicity in the electronic structure \cite{Pletikosic2009,Starodub2011} and buckling of the carbon layer \cite{NDiaye2006,Borca2010} depends on the strength and local variation of the interaction between graphene and substrate, which ranges from strong chemisorption to weak physisorption \cite{Preobrajenski2008}. A key parameter for this strength is the height $h$ of the carbon adsorbate, which is, however, difficult to assess by experiment: Analysis by x-ray or electron diffraction necessitates a large number of fitting parameters because of extended unit cells \cite{Moritz2010,Martoccia2010}. In addition, the x-ray scattering amplitude of C is low \cite{Martoccia2010}. Determination of $h$ by scanning tunneling microscopy (STM) and atomic force microscopy is also questionable \cite{STMAFMcomment}. Only for a few systems such measurements have been performed, and consensus has not been reached [see \cite{Borca2010} for Ru(0001)], except for highly commensurate graphene. 

In density functional theory (DFT) a quantitative description of the interaction between graphene and metal is a challenge, because the most commonly used exchange-correlation functionals [local density approximation (LDA) and generalized gradient approximation (GGA)] are (semi)local \cite{LDAGGAcomment}. They lack the nonlocal-correlation effects responsible for van der Waals (vdW) interaction \cite{Vanin2010,Brako2010}. As an example, for the weakly bound system graphene/Ir(111) investigated here, DFT-GGA calculations  result in a mean value of $\bar{h} \approx 3.9$~{\AA} \cite{NDiaye2006} and very low binding energies $E_{\rm b}$ of only a few meV per C atom \cite{Lacovig2009,NDiaye2006}, thus contradicting the experimentally observed formation and stability of graphene at about 1300~K \cite{Coraux2008}. Such a low $E_{\rm b}$ is also not plausible in view of the observation that the peeling force needed to remove graphene from Ir(111) \cite{Herbig2011} is higher than what is necessary to exfoliate graphite (experimentally determined $E_{\rm b} = -52$\,meV per C atom \cite{Zacharia2004}). LDA calculations \cite{Feibelman2008,Feibelman2009} succeed in binding graphene in accord with $\bar{h} = 3.42$~{\AA}. However, conceptually this success is unsatisfying and questionable as the error of disregarding the vdW interaction is diminished by error cancellation: The LDA has a well known tendency to overbind. Only recently, the first truly nonlocal correlation functional vdW-DF \cite{Dion2004} was developed and successfully applied to simple vdW-bonded systems~\cite{Langreth2009}, thereby opening a new perspective to correctly describe also complex systems with significant vdW bonding.   

Without a proper understanding of binding between graphene and Ir(111), also other discrepancies will be hard to resolve: Whereas previously the absence of any interaction between the Dirac cone and the Ir 5d bands was assumed \cite{Pletikosic2009}, more recent experiments point towards a hybridization close to the Fermi level \cite{Starodub2011}. Such a hybridization also influences the properties of phonons in graphene: Strikingly, both the presence \cite{VoVan2011} and the absence \cite{Starodub2011} of the respective Raman modes have been reported. Even the commonly assumed shape of graphene \cite{NDiaye2006} appears inverted in a recent AFM experiment \cite{Sun2011}.

In this Letter we further the understanding of graphene on metal by (i) performing an x-ray standing wave (XSW) experiment, which allows us to unambiguously specify the average bond distance $\bar{h}$ and the amplitude of buckling $\Delta h$ and (ii) applying nonlocal vdW-DF using a realistically large supercell. The comparison of experiment and theory allows to benchmark the functionals used to describe bonding of graphene and the bonding of $\pi$-conjugated systems on metals in general. 

The experiments were performed at the ESRF undulator beamline ID32 \cite{Zegenhagen2010}. Ir(111) (mosaicity $0.2^{\circ}$ determined from x-ray diffraction rocking curves) was cleaned in UHV by cycles of $1.5$~keV Ar$^+$ bombardment and annealing to $1420~$K. Graphene was grown by repeated cycles of room temperature C$_2$H$_4$ adsorption followed by thermal decomposition at $1420$~K (temperature programmed growth or TPG). Well oriented graphene forms, leading to the characteristic moir{\'e} pattern in low energy electron diffraction \cite{Pletikosic2009}. We prepared a coverage of $(0.39 \pm 0.03)$~ML via two TPG cycles and of $(0.63 \pm 0.04)$~ML via four cycles. For coverage estimation we used the fact that a fraction of $(0.22 \pm 0.02)$ of the free Ir(111) surface is covered with graphene after each TPG cycle \cite{Coraux2009}. Photoelectron spectra (overall resolution $400$~meV) were recorded using a hemispherical electron analyzer. The reflectivity of the sample was measured by directing the reflected beam to an insulated metal plate and measuring the resulting electron emission current. Scanning tunneling microscopy (STM) was performed on another Ir(111) sample in a separate UHV system where the growth was repeated \cite{XSW_supplement}.

An X-ray standing wave was created in the interface region of a crystal using Bragg reflection. We used the (111) reflection at an angle close to $90^{\circ}$ at $2.801$~keV. The XSW maxima, which are periodic with the Ir(111) lattice planes, are shifted by half the lattice plane distance when scanning through the Bragg reflection by changing the beam energy. The photoelectron yield of adsorbates during such a scan depends on their height above the surface. From the results of an XSW measurement, one obtains the coherent position $P^{\rm H}$ and the coherent fraction $F^{\rm H}$ \cite{Vartanyants2000}, which roughly correspond to the mean adsorbate height and the spread around this value \cite{XSW_supplement}.

\begin{figure}[hbt]
\begin{center}
\includegraphics[width=8.2cm]{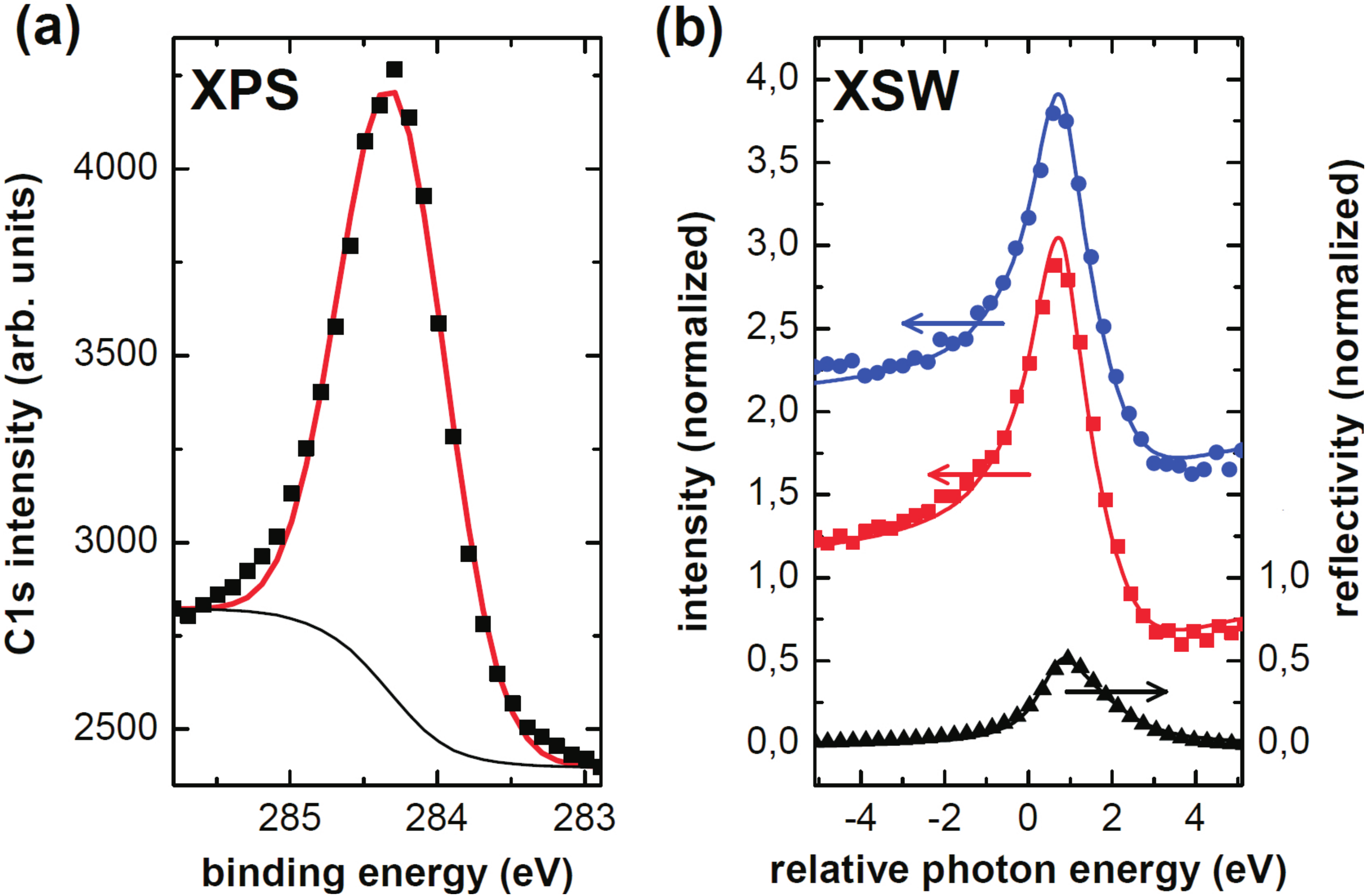}
\end{center}
\caption{(Color online) (a) Squares: X-ray photoelectron intensity of the C1s-peak for $0.63$~ML of graphene on Ir(111). Thin line: Shirley-type background, Thick line: Gaussian fit (full width at half maximum 0.85~eV) above background. (b) X-ray reflectivity (triangles) normalized using the fit to the data (solid line \cite{Zegenhagen1993}) and exemplary C1s photoemission yield normalized to an off-Bragg yield of 1 (squares: $0.39$~ML, circles: $0.63$~ML, shifted upwards by unit of $1$ for easier readability). Solid lines: fits to the data \cite{Zegenhagen1993,Zegenhagen2010} taking into account additional broadening due to the mosaicity of the sample and the bandpass of the X-ray beam. \label{fig:XSW}}
\end{figure}

From the photoelectron spectra [compare Fig.~\ref{fig:XSW}(a)] we determined the C1s binding energy to $(284.2 \pm 0.1)$~eV, in agreement with Ref.~\onlinecite{Preobrajenski2008}. After subtraction of a Shirley background, all C1s spectra could be fitted well with a single Gaussian. From the resulting peak area as a function of photon energy [Fig.~\ref{fig:XSW}(b)] and averaging over several scans for the same preparation, the structural parameters $P^H=0.53 \pm 0.01$ and $F^H=0.87 \pm 0.04$ for $0.39$~ML and $P^H=0.52 \pm 0.01$ and $F^H=0.74 \pm 0.04$ for $0.63$~ML are obtained \cite{Zegenhagen2010}. To interpret these values, we tested three simple height distribution functions (Gaussian, rectangular, p6m layer with Fourier components up to first order \cite{XSW_supplement}). It turns out that almost the same structural parameters give the best fits to $P^H$ and $F^H$ for all three models, showing the robustness of our interpretation. We determined $\bar{h}=(3.38 \pm 0.04)$~{\AA} \cite{XSW_supplement} for both coverages, and a standard deviation of $\sigma_{h}=(0.19 \pm 0.03)$~{\AA} for 0.39~ML and $\sigma_{h}=(0.27 \pm 0.04)$~{\AA} for 0.63~ML. Here the errors contain both the experimental uncertainty as well as the small deviations resulting from the choice of the model. The standard deviation gives an upper limit for the possible corrugation of graphene. The measured mean height is similar to the interlayer distance in graphite of $3.36$~\AA. This already indicates that only weak bonding can be at place. Where applicable, the standard deviation translates to $\Delta h=(0.6 \pm 0.1)$~{\AA} for 0.39~ML and $\Delta h = (1.0 \pm 0.2)$~{\AA} for 0.63~ML, see discussion below.

\begin{figure*}[htb]
\begin{center}
\includegraphics[width=16cm]{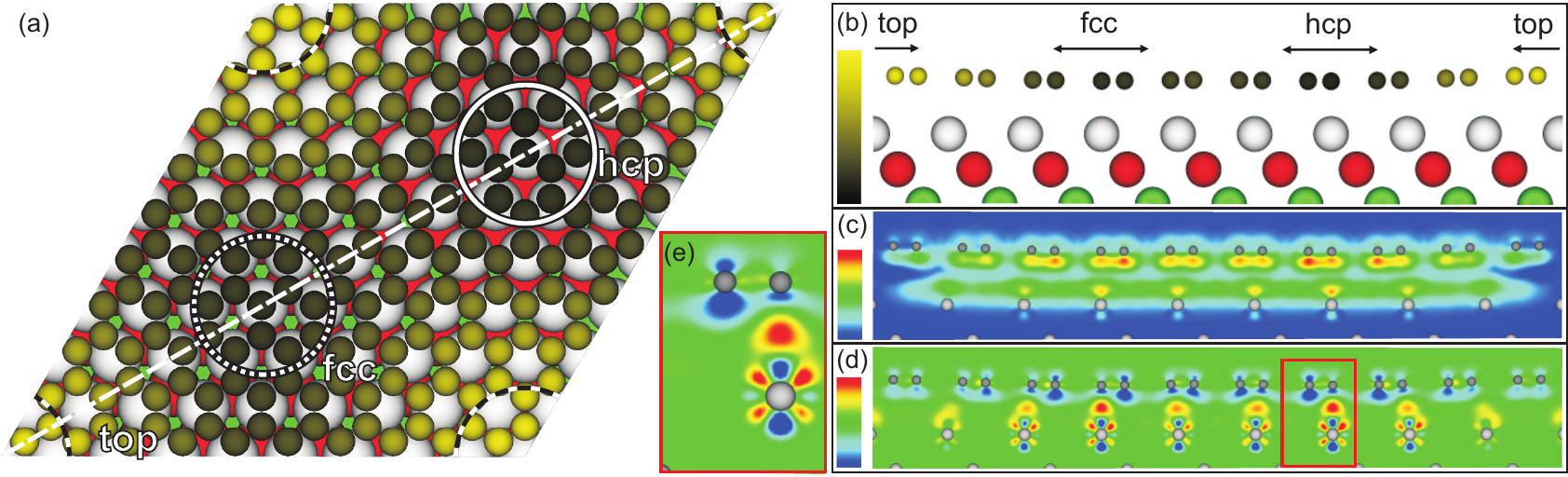}
\end{center}
\caption{(Color online) (a) Top view and (b) side view [cut along the dashed line in (a)] of the relaxed structure of graphene/Ir(111) obtained by DFT including vdW interactions. Regions of high-symmetry stacking (fcc, hcp, top) are marked by circles (a) or arrows (b-d). The color scale in (b) and (a) ranges from $h=3.20$~{\AA} (dark) to $h=3.65$~{\AA} (yellow). (c) Visualization of the nonlocal correlation binding-energy density $e_{\mathrm{c}}^{\mathrm{nl,bind}}(\textbf{\textit{r}})$ caused by adsorption. The color scale ranges from $e_{\mathrm{c}}^{\mathrm{nl,bind}}(\textbf{\textit{r}}) = 0$~meV {\AA}$^{-3}$ (blue) to $e_{\mathrm{c}}^{\mathrm{nl,bind}}(\textbf{\textit{r}}) = -28.3$~meV {\AA}$^{-3}$ (red). (d) Charge transfer upon adsorption. The color scale ranges from $\Delta \rho = -0.0138$~e {\AA}$^{-3}$ (blue) to $\Delta \rho = 0.0138$~e {\AA}$^{-3}$ (red). A negative value indicates loss of electron density. (e) Magnified view of red box in (d) Views from different angles can be obtained in \cite{XSW_supplement}. \label{fig:DFT_results}}
\end{figure*}

{\emph{Ab initio}} DFT calculations have been performed by using the projector augmented wave method \cite{Blochl1994} with the PBE-GGA functional \cite{Perdew1996} as implemented in VASP \cite{Kresse1993}. To obtain a reliably relaxed adsorbate-surface geometry~\cite{Atodiresei2009} we have implemented vdW forces using the semi-empirical method DFT-D \cite{Grimme2006}, where the Ir $C_6$ coefficient was determined by comparing DFT-D and nonlocal vdW-DF calculations for several adsorption geometries of benzene on Ir(111). For the final relaxed graphene-Ir(111) geometry, the total energy was calculated with vdW-DF in a post-processing approach using the JuNoLo code \cite{Lazic2010}. In this method,  the vdW-DF total energy is evaluated with the electron density obtained from the GGA calculations and its value is not changed when performing a full selfconsistent vdW-DF calculation~\cite{Thonhauser2007}.

The system was modeled by $(10 \times 10)$ unit cells of graphene on $(9 \times 9)$ cells of Ir(111) resulting in a ratio of the surface unit cell lengths of $a_{\rm C}/a_{\rm Ir} = 1.111$, close to the experimental value of $a_{\rm C}/a_{\rm Ir} = 1.107$ for flakes with an average size of 1000~{\AA}. We used a slab geometry including in total about 600 atoms (4 layers of Ir and 1 layer graphene) in the unit cell. In previous studies for graphene on metals the vdW-DF functional was applied using much smaller unit cells \cite{Vanin2010,Brako2010,Hamada2010} where effects of the moiré superstructure are suppressed. Plane waves with a kinetic energy up to $400$~eV have been included in the basis set and the Brillouin zone was sampled by a $3\times3\times1$ $k$-mesh. Geometry optimization was obtained by relaxing the top 2 Ir layers and the graphene layer including the long--range vdW forces as described above with a force threshold set to $1$~meV/\AA.

The geometry resulting from this calculation is shown in Fig.~\ref{fig:DFT_results}(a). The largest height of $3.62$~{\AA} is found in regions where the center of the C hexagon is located on top of an Ir atom (top region) and the lowest height of $3.27$~{\AA} in the hcp-region (center of C hexagon above threefold coordinated hcp site), slightly lower than the one of $3.29$~{\AA} in the fcc-region. The mean height is $\bar{h}=3.41$~{\AA}, in excellent agreement with the experimental value. The corrugation resulting from DFT including vdW is $\Delta h=0.35$~{\AA} or $\sigma_{h}=0.09$~{\AA}, and thus safely within the upper limits determined by experiment for both coverages.

The averaged binding energy per C atom in our calculations is $E_{\rm b}=-50$~meV/C. The vdW-DF approach makes it possible to distinguish between local and non-local contributions to the overall binding energy \cite{XSW_supplement}. The nonlocal correlation energy $E_{\mathrm{c}}^{\mathrm{nl}}$ can be expressed as a function of the charge density $\rho(\textbf{\textit{r}})$

\begin{equation*}
E_{\mathrm{c}}^{\mathrm{nl}}= \frac{1}{2} \iint
  d\textbf{\textit{r}} d\textbf{\textit{r}}'
  \rho(\textbf{\textit{r}}) \phi(\textbf{\textit{r}},\textbf{\textit{r}}')
  \rho(\textbf{\textit{r}}')
  =  \int d\textbf{\textit{r}}
          e_{\mathrm{c}}^{\mathrm{nl}}(\textbf{\textit{r}})\, ,
\label{eq:NonLocal}
\end{equation*}
where $\phi(\textbf{\textit{r}},\textbf{\textit{r}}')$  is the kernel function (see discussion in Ref.~\onlinecite{Dion2004}) and $e_{\mathrm{c}}^{\mathrm{nl}}(\textbf{\textit{r}})$ is the nonlocal correlation energy density at each point $\textbf{\textit{r}}$ in real space.

The distribution of the nonlocal correlation binding energy density $e_{\mathrm{c}}^{\mathrm{nl,bind}}(\textbf{\textit{r}})$ (i.e.\ the change in $e_{\mathrm{c}}^{\mathrm{nl}}(\textbf{\textit{r}})$ caused by adsorption \cite{XSW_supplement}) is shown in Fig.~\ref{fig:DFT_results} (c). Note that $e_{\mathrm{c}}^{\mathrm{nl,bind}}(\textbf{\textit{r}})$ arises from a non-local quantity as its value at a given point depends on the interaction of the charge density at this point with the one at all other points. The fact that this energy density is broadly distributed in a layer just above the metal surface and just below the graphene plane (note that for graphene we cut alternately through bonds and hexagon centers) shows that the polarization effects responsible for the vdW interaction are spread over the entire sheet, which is a clear fingerprint of a vdW-bonded $\pi$-conjugated system~\cite{Atodiresei2009,Vanin2010}. 

For the relaxed geometry shown in Fig.~\ref{fig:DFT_results}(a), the binding energy $E_{\rm b}^{\rm GGA}$ calculated in GGA is repulsive ($\approx +20$~meV/C) while the non-local binding energy contribution $E_{\rm b}^{\rm nl} \approx -70$~meV/C  is attractive summing up to the total vdW-DF binding energy $E_{\rm b}=-50$~meV/C~\cite{XSW_supplement}.  Contrary to the impression given by these averaged values, the binding is not pure physisorption, but chemically modulated. This becomes obvious when analyzing the charge transfer caused by adsorption [Figs.~\ref{fig:DFT_results}(d) and (e)]. In the hcp- and fcc-regions a small charge transfer from graphene towards the substrate takes place. A C atom sitting directly atop of an Ir atom [see Fig.~\ref{fig:DFT_results}(e)] hybridizes its C(2p$_z$) orbital with the Ir(5d$_{3z^2 - r^2}$) orbital. As a result charge accumulates just in between the C atom and the Ir atom, indicating formation of a weak covalent bond. This charge is provided primarily from the neighbor C atoms to the bondforming ones sitting atop of Ir atoms. The charge deficit of these neighbors explains their tendency to bind additionally deposited metal atoms \cite{NDiaye2006}. 
This charge transfer is intimately related to the nonlocal binding energy density [Fig.~\ref{fig:DFT_results}(c)] localized in specific regions close to those Ir atoms where the charge transfer from graphene to substrate occurs. This indicates those sites at the metal substrate that become more polarizable upon adsorption due to the graphene-surface interaction.  In total, graphene has lost $\approx 0.01$~electrons/C of charge resulting in slight p-doping.

A by-product of DFT calculations are the Kohn-Sham eigenstates. Due to the large supercell in our calculations the calculated bands are multiply folded. To extract the dispersion relations for the entire Brillouin zone heavy postprocessing would be necessary \cite{Ku2010}. Nevertheless, taking only the states with a large projection on the carbon atoms, thereby filtering out the substrate states, we identify the Dirac cone of the adsorbed graphene in the vicinity of the Fermi level. The Dirac point is shifted $0.2$~eV above $E_{\rm Fermi}$ consistent with the experimental value of $0.1$~eV \cite{Pletikosic2009}.

\begin{figure}[hbt]
\begin{center}
\includegraphics[width=6cm]{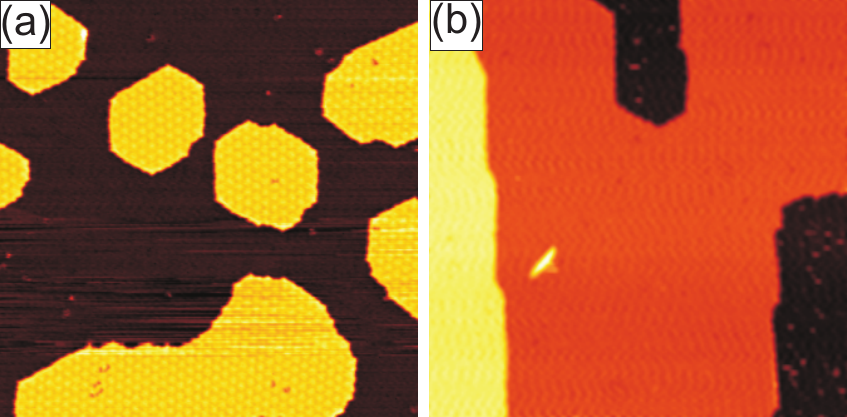}
\end{center}
\caption{(Color online) STM topographs of graphene on Ir(111). (a) 0.39 ML. Flakes with the characteristic graphene moir\'{e} are visible. (b) $0.63$~ML. A sheet extends over a wide area including a substrate step. Lowest regions are bare Ir(111). Image widths $1000$~\AA, bias voltage $U_{\rm sample}=-1$~V, tunneling current $I=0.2$~nA (a), $I=0.06$~nA (b). \label{fig:STM}}
\end{figure}

Finally we address the experimentally observed coverage dependence of the corrugation. To obtain insight into this dependence we analyzed the graphene films using STM (Fig.~\ref{fig:STM}). For $0.39$~ML ($\Delta h = 0.6$~\AA), graphene is present in the form of flakes with a mean size of $\approx 500$~\AA, most of them located on extended terraces. For $0.63$~ML ($\Delta h = 1.0$~\AA) large coalesced flake agglomerates form with linear dimensions above $1000$~{\AA}. Here, the graphene overgrows substrate steps \cite{Coraux2008} and occasionally wrinkles are found \cite{NDiaye2009b}, which are absent for the smaller flakes. We speculate that in the different geometries the shrinking of the substrate while cooling down from the high growth temperatures \cite{NDiaye2009b} has distinct effects: As small flakes are able to float they remain relaxed and flat. Larger flakes pinned to and overgrowing steps are unable to float. One mechanism to avoid build up of stress is then to buckle. Dedicated experiments to substantiate these speculations are necessary and under way. Such a strain dependence of the corrugation is out of reach in our simulations, as in the vdW-DF calculations the $10 \times 10$ graphene sheet matches the $9 \times 9$ substrate mesh without strain.


\begin{acknowledgments}
P. L.\ thanks the AvH-foundation for a research fellowship. Financial support of the DFG (Forschergruppe 912, Bu2197/2-1, Mi581/17-2) is gratefully acknowledged. The computations were performed on JUROPA and JUGENE supercomputers at the J\"ulich Supercomputing Center, Forschungszentrum J\"ulich (Germany). The authors thank the staff at ESRF for technical assistance. \end{acknowledgments}

\newpage

\onecolumngrid

\renewcommand{\baselinestretch}{1.50}\normalsize

\section{Graphene on Ir(111): Physisorption versus chemical repulsion - Supplementary materials}

\subsection{Scanning tunneling microscopy}

The STM measurements depicted in Fig. 3 of our manuscript have been performed in a UHV system at the University of Cologne using a home-build beetle-type STM (the system is described in Detail in Refs.~\onlinecite{PhD_Michely,PhD_Busse}). For the measurements we used an Ir(111) crystal from the same manufacturer with identical specifications and prepared the crystal and the graphene using the same experimental parameters as for the XSW experiments. 
\subsection{Analysis of the X-ray standing waves data}

Since we detected the differential photoelectron yield, multipole contributions had to be taken into account and the data was fitted according to \cite{Vartanyants2000}. Strictly, XSW can only give you the relative height with respect to the crystal lattice planes, i.e. $\bar{h}$ mod $d_{\rm Ir(111)}$. Here we disregarded unphysically small or large values. To interpret the experimentally determined parameters $P^H$ and $F^H$, we tested different height distribution functions $G(z)$: A Gaussian function:

\begin{equation}
G(z)=\frac{1}{\sigma \sqrt{2 \pi}} e^{-\frac{1}{2}\left(\frac{z - z_0}{\sigma}\right)^2},
\end{equation}

a rectangular function with width $s$ and height $1/s$ (note that for such a distribution the standard deviation is given as $\sigma = \sqrt{\frac{1}{12}}s$)

\begin{equation}
G(z) =  \begin{cases}
        \frac{1}{s} & {\rm if} |z - z_0| \leq \frac{s}{2}, \\
         0 &  {\rm if} |z - z_0| > \frac{s}{2},
        \end{cases}
\end{equation}

and the height distribution of a sixfold symmetric (p6m) layer described by a superposition of three cosine waves (eggbox-model), i.e. containing only Fourier coefficients up to first order:

\begin{equation}
h(\vec{r})=\bar{h} + \frac{2}{9} \Delta h \left(\cos(\vec{k}_1  \vec{r})+\cos(\vec{k}_2 \vec{r})+\cos(\vec{k}_3 \vec{r}) \right)
\end{equation}

with  $|\vec{k}_1| =|\vec{k}_2| =|\vec{k}_3| $ and $\angle \vec{k}_1, \vec{k}_2 =  \angle \vec{k}_2, \vec{k}_3=120^{\circ}$,

The eggbox-model is motivated by the appearance of C/Ir(111) in STM.  Note that the exact shape resulting from the calculations can well be fitted with this eggbox-model with a root mean square deviation of only 0.006~{\AA}/C. When in the studies for graphene on Ru(0001) cited in the main manuscript only the highest and lowest value of $h$ were explicitly given, we estimated $\bar{h}$ on the basis of the eggbox-model.

\subsection{Details of the calculation}

The vdW-DF approach makes it possible to distinguish between local and non-local contributions to the overall binding energy. Technically, the correlation part of the generalized gradient approximation (GGA) functional is replaced by a sum of the local density approximation (LDA) correlation functional $E_{\mathrm{LDA,c}}$ and a nonlocal term $E_{\mathrm{c}}^{\mathrm{nl}}$ such that the total energy is given by

\begin{equation}
E_{\mathrm{vdW-DF}}=E_{\mathrm{GGA}}-E_{\mathrm{GGA,c}}+E_{\mathrm{LDA,c}}+E_{\mathrm{c}}^{\mathrm{nl}},
\label{eq:vdW-DF}
\end{equation}
where $E_{\mathrm{GGA}}$ is the self-consistent GGA total energy evaluated in a conventional DFT calculation. The nonlocal correlation energy $E_{\mathrm{c}}^{\mathrm{nl}}$ can be expressed as a function of the charge density $n(\textbf{\textit{r}})$,

\begin{equation}
E_{\mathrm{c}}^{\mathrm{nl}}= \frac{1}{2} \int \int
  d\textbf{\textit{r}} d\textbf{\textit{r}}'
  n(\textbf{\textit{r}}) \phi(\textbf{\textit{r}},\textbf{\textit{r}}')
  n(\textbf{\textit{r}}'),
\label{eq:NonLocal}
\end{equation}
where the kernel function $\phi(\textbf{\textit{r}},\textbf{\textit{r}}')$ is discussed in detail in Ref.~\onlinecite{Dion2004}.

To evaluate Eq.~\ref{eq:NonLocal}, it can be rewritten as

\begin{align}
\begin{split}
E_{\mathrm{c}}^{\mathrm{nl}}
  &= \frac{1}{2} \int \int
     d\textbf{\textit{r}} d\textbf{\textit{r}}'
     n(\textbf{\textit{r}}) \phi(\textbf{\textit{r}},\textbf{\textit{r}}')
     n(\textbf{\textit{r}}'), \\
  &= \int d\textbf{\textit{r}} n(\textbf{\textit{r}})
          \left[ \frac{1}{2} \int d\textbf{\textit{r}}'
	         \phi(\textbf{\textit{r}},\textbf{\textit{r}}')
	       n(\textbf{\textit{r}}')
	   \right], \\
  &= \int d\textbf{\textit{r}} n(\textbf{\textit{r}})
          \epsilon_{\mathrm{c}}^{\mathrm{nl}}(\textbf{\textit{r}}), \\
  &= \int d\textbf{\textit{r}}
          e_{\mathrm{c}}^{\mathrm{nl}}(\textbf{\textit{r}}),
\label{eq:NonLocal_RealSpace}
\end{split}
\end{align}
where $e_{\mathrm{c}}^{\mathrm{nl}}(\textbf{\textit{r}})$ is the nonlocal correlation energy density for each point $\textbf{\textit{r}}$ in the real space.

Then, the nonlocal correlation binding energy density $e_{\mathrm{c}}^{\mathrm{nl,bind}}(\textbf{\textit{r}})$ plotted in Fig. 3 is  defined as

\begin{equation}
e_{\mathrm{c}}^{\mathrm{nl,bind}}(\textbf{\textit{r}})=
e_{\mathrm{c}}^{\mathrm{nl,sys}}(\textbf{\textit{r}})-
(e_{\mathrm{c}}^{\mathrm{nl,graphene}}(\textbf{\textit{r}}) +
 e_{\mathrm{c}}^{\mathrm{nl,Ir(111)}}(\textbf{\textit{r}}) ),
\end{equation}
where the $e_{\mathrm{c}}^{\mathrm{nl,sys}}(\textbf{\textit{r}})$ denotes the nonlocal correlation energy density of the graphene--Ir(111) system, while $e_{\mathrm{c}}^{\mathrm{nl,graphene}}(\textbf{\textit{r}})$ and $e_{\mathrm{c}}^{\mathrm{nl,Ir(111)}}(\textbf{\textit{r}})$ are the nonlocal correlation energy density at point \textbf{\textit{r}} of the graphene and Ir(111) in their graphene--surface relaxed geometry, respectively. The non-local binding energy contribution, $E_{\rm b}^{\rm nl}$, per C atom, is given by the integal of $e_{\mathrm{c}}^{\mathrm{nl,bind}}$ over the space of the unit cell, plus a small contribution arising from  the binding
energy differences of $E_{\mathrm{GGA,c}}+E_{\mathrm{LDA,c}}$ in Eq.~\ref{eq:vdW-DF}, divided by the number of C in the unit cell.


\begin{thebibliography}{10}
\bibitem{LDAGGAcomment}
In LDA, the functional depends on the charge density $\rho(\mathbf{r})$,
  whereas in GGA it also depends on $\vec{\nabla} \rho(\mathbf{r})$.
\end{thebibliography}
\end{document}